\documentstyle[11pt]{article}
\begin{document}
{\setlength{\oddsidemargin}{1.2in}
\setlength{\evensidemargin}{1.2in} } \baselineskip 0.55cm
\begin{center}
{\bf Modified Hawking radiation of stationary and nonstationary Kerr-Newman-de Sitter black hole}
\end{center}
\begin{center}
{S. Christina}$^{1}$,\,\,\,\,{\rm T.\, Ibungochouba\,Singh$^{1*}$}
\end{center}
\begin{center}
1. Department of Mathematics, Manipur University, Canchipur, 795003, India\\

\end{center}
\date{}
\begin{center}
 * E-mail: ibungochouba@rediffmail.com
\end{center}
\begin{abstract}
In this paper, the tunneling of vector boson particles across the event horizon of Kerr-Newman-de Sitter (KNdS) black hole is investigated using the WKB approximation and Hamilton-Jacobi ansatz to Proca equation. The tunneling probability and radiation spectrum at the event horizon of KNdS black hole is obtained. The modified Hawking radiation of stationary and non-stationary KNdS black hole are also investigated using the Rarita-Schwinger ansatz to the deformed Hamilton-Jacobi method and new tortoise coordinate transformation. It is observed that the modified Hawking temperatures near the event horizon for both stationary and nonstationary space times are dependent on the black hole parameters.
\end{abstract}
\section{Introduction}
The black hole formation is a robust prediction in general relativity and the characteristics of black holes have been studied based on general relativity [1]. Hawking [2, 3] investigated the thermodynamics  of black hole by combining general relativity with quantum filed theory which indicates that black hole radiates like a black body radiation. Since then, the relationship between the entropy and horizon area was developed successfully [4, 5]. The Hawking radiation of black hole is also investigated by using the radial null geodesic method [6]. In this method, the potential barrier is produced by the outgoing particles and the imaginary part of the radial action  is calculated by applying WKB approximation. The dynamical equations of the spin-1/2 and spin-3/2 fermions were described by Dirac equation and Rarita-Schwinger equation in curved spaced times respectively. The tunneling of spin-1/2 fermions for the charged black hole using the semiclassical methods has been investigated in [7, 8]. The quantum tunneling of boson and fermions using Hamilton-Jacobi method have been studied in [9-11]. It is shown that the characteristics of both bosons and fermions may be described by one single equation--Hamilton-Jacobi equation. The Hawking radiation as tunneling beyond the semiclassical approximation has been investigated by using Hamilton-Jacobi method. The correction to Bekenstein-Hawking entropy for the Schwartzschild black hole and Kerr black hole have been derived by applying Hamilton-Jacobi equation and the first law of black hole thermodynamics [12-14]. Many fruitful results have been obtained in [15-17] by using Hamilton-Jacobi method. Refs. [18-20] investigated the  information loss and the tunneling radiation of black holes. 

Using Klein-Gordon equation, Chen and Huang [21, 22] studied the tuneling of static and  dynamic black hole. It is learnt that the dynamical equations of spin-1/2 and spin-3/2 fermions in the curved spacetime are governed by a matrix equation which is very complicated to begin the research process. To investigate the dynamical features of particles in curved spacetime, most of researchers used Hamilton-Jacobi equation. It is known that the process of studying the dynamical characteristics of particles, specially the tunneling of fermions with non-zero spin in curved spacetime is simplified by Hamilton-Jacobi method.
 The nonstationary black hole will be the actual black hole in the universe, so that the information conservation and the thermodynamics properties of nonstationary black holes and violent merger of black holes are needed to be studied in depth.
 
 The quantum field theory implies that Lorentz symmetry could be modified at high energy. The basic relation in both general theory of relativity and quantum field theory is the relativistic dispersion relation $E^2=m^2c^4+p^2c^2$ but it has not been fully established yet. The generally accepted scale of this correction term is approximately equal to the Planck scale. 
  The tunneling of vector particles from the black hole by applying the Hamilton-Jacobi ansatz to Proca equation and WKB approximation was proposed by Kruglov [23, 24]. The Hawking temperature corresponding to scalar particle emission is same as the emission temperature of Schwarzschild black hole. Applying Proca equation, many interesting results have been obtained in [25-30]. Refs. [31, 32] proposed the tortoise coordinate transformation to study the Hawking radiation of stationary and nonstationary black hole, in which gravitational field is assumed to be independent on time.  Chandrasekhar [33] and Bonner and Vaidya [34] showed that the radial equation and angular equation of Dirac particle expressed in Newman-Penrose formalism can be separated for stationary rotating or nonrotating  uncharged black hole.
 
 In this paper, we will discuss the quantum tunneling process of massive vector boson particles near the event horizon of static and stationary black hole by applying Proca equation, WKB approximation and Feynman prescription. The radiation spectrum of the emitted vector boson particle which is related to Boltzman factor of emission in accordance with semiclassical approximation for the static and stationary KNdS black hole will be obtained. The modified Hawking temperatures near the event horizon of stationary and nonstationary KNdS black hole will also be investigated by using the Rarita-Schwinger ansatz to the deformed Hamilton-Jacobi equation, Feynman Prescription and tortoise coordinate transformation. We will show that different methods lead to the different modified Hawking temperatures in the nonstationary black hole. In the absence of deformation term, these four different methods will lead to the same Hawking temperatures for stationary KNdS space time.

 The paper is organised as follows. In sec. 2, we discuss the Hawking temperature of Kerr-Newman-de Sitter (KNdS) black hole in the static coordinate system using Hamilton-Jacobi ansatz to Proca equation and WKB approximation. In sec. 3, the deformed Hamilton-Jacobi equation is derived from Rarita-Schwinger equation and WKB approximation. In sec. 4, the modified Hawking temperature near the event horizon of KNdS black hole is derived by using deformed Hamilton-Jacobi equation. In sec. 5, the modified Hawking temperature of nonstationary KNdS black hole is studied by using deformed Hamilton-Jacobi equation and tortoise coordinate transformation. Discussion and conclusion are given in sec. 6 and sec. 7 respectively.
\section{Kerr-Newmann-de Sitter black hole}
The line element of Kerr-Newman-de Sitter black hole in Boyer-Lindquist coordinate system   [35] can be written as 
\begin{eqnarray}
ds^2&=&-\frac{\Delta-\Delta_\theta a^2\sin^2\theta}{\rho^2 \Xi^2}dt^2+\frac{\rho^2}{\Delta}dr^2+\frac{\rho^2}{\Delta_\theta}d\theta^2-\frac{2a[\Delta_\theta(r^2+a^2)-\Delta]\sin^2\theta}{\rho^2 \Xi^2}dt d\varphi \cr &&+\frac{\Delta_\theta(r^2+a^2)^2-\Delta a^2\sin^2\theta}{\rho^2\Xi^2}\sin^2\theta d\varphi^2,
\end{eqnarray}

 where the symbols $\rho$, $\Delta$, $\Delta_\theta$ and \,$\Xi$ are defined respectively by
\begin{eqnarray}
\rho^2 &=& r^2+a^2\cos^2\theta, \,\,\,\,\,\,\, \Xi=1+\frac{\Lambda a^2}{3}, \,\,\,\,\, 
\Delta_{\theta} = \Big(1+\frac{\Lambda a^2\cos^2\theta}{3}\Big),\cr 
\Delta &=& \Big(1-\frac{\Lambda r^2}{3}\Big)(r^2+a^2)-2Mr+Q^2.
\end{eqnarray}
Here Eq. (1) represents KNdS black hole for $\Lambda>0$ and  anti-KNdS black hole for
 $ \Lambda<0 $. $ M $ is the mass of black hole and $ a $ is the rotational parameter. If $\frac{1}{\Lambda}>>M^2>a^2+Q^2$, then $ \Delta=0 $ gives the four real roots namely $ r_c,\ r_h,\ r_{+} $ and $r_{-}$ $(r_c>r_h>r_{+}>0>r_{-})$. The largest root $r_c$ represents the location of the cosmological horizon, $r_h$ corresponds to the location of the event horizon and $r_{+}$ indicates the location of Cauchy horizon. The negative root $r_{-}$ represents the another Cosmological horizon on the other side of ring singularity at $r=0$ and another infinity [36]. To investigate the original and modified Hawking radiation near the event horizon of KNdS black hole, $\Delta$ can be factorised as [37]

\begin{eqnarray}
\Delta=(r-r_h)\Delta^{\prime}(r_h).
\end{eqnarray}
The contravariant metric components of the Kerr-Newman-de Sitter black hole are
\begin{eqnarray}
g^{00}&=&-\frac{[\Delta-\theta(r^2+a^2)^2-\Delta a^2 \sin^2\theta]\Xi^2}{\Delta \Delta_\theta \rho^2}, \cr
g^{11}&=&\frac{\Delta}{\rho^2}, \,\,\,\,
 g^{22}=\frac{\Delta_\theta}{\rho^2},\cr 
 g^{03}&=&g^{30}=-\frac{a\Xi^2[\Delta_\theta(r^2+a^2)-\Delta]}{\Delta \Delta_\theta \rho^2},\cr 
 g^{33}&=&\frac{[\Delta-\Delta_\theta a^2 \sin^2\theta]\Xi^2}{\rho^2\Delta\Delta_\theta \sin^2\theta}.
\end{eqnarray}
Let $\phi=\varphi-\Omega t$  and\,\,\, $\Omega=-\frac{g_{14}}{g_{44}}$. Then Eq.(1) becomes
\begin{eqnarray}
ds^2&=&-\frac{\Delta \Delta_\theta \rho^2}{ \Xi^2 [\Delta_\theta(r^2+a^2)^2-\Delta a^2 \sin^2\theta]}dt^2 + \frac{\rho^2}{\Delta}dr^2 +\frac{\rho^2}{\Delta_\theta}d\theta^2\cr && +\frac{[\Delta_\theta(r^2+a^2)^2-\Delta a^2 \sin^2\theta]\sin^2\theta}{\rho^2 \Xi^2}d\phi^2,
\end{eqnarray}
where the angular velocity at the event horizon is given by
\begin{eqnarray}
\Omega=\frac{a}{r^2_h+a^2}.
\end{eqnarray}
From Eq. (5), the surface gravity  and the Hawking temperature near the event horizon $ r=r_{h} $ of stationary KNdS black hole are defined  by 26, 38]
\begin{eqnarray}
\kappa &=& \lim_{g_{00}\rightarrow 0}\Bigg(-\frac{1}{2}\sqrt{-\frac{g^{11}}{g_{00}}} \frac{d g_{00}}{dr}\Bigg)\cr
&=& \frac{r_{h} -M - \frac{2}{3}\Lambda r_{h}^{3} - \frac{1}{3}\Lambda r_{h}a^{2}}{\Xi(r_{h}^{2} + a^{2})}.
\end{eqnarray}
and
\begin{eqnarray}
T_{H} &=& \frac{\kappa}{2\pi}\cr
&=& \frac{1}{2\pi}\Bigg[\frac{r_h-M-\frac{2}{3}\Lambda r_h^3-\frac{\Lambda}{3}r_h a^2}{ \Xi (r_h^2+a^2)}\Bigg].
\end{eqnarray}
We study the tunneling of the massive vector particle near the event horizon of KNdS black hole. The wave function $ \Psi $ in the semiclassical approximation satisfies the Proca equation as

\begin{equation}
\frac{1}{\sqrt{-g}}\partial_{_{\mu}}(\sqrt{-g}\Psi^{\mu \nu})+\frac{m^2}{{\hbar}^{2}}\Psi^\nu=0,
\end{equation}
where $\Psi_{\mu \nu}=\partial_{\mu}\Psi_{\nu}-\partial_{\nu}\Psi_{\mu}$ and $\Psi^{\mu \nu}$ is an anti-symmetric tensor. $ m $ and $ \hbar $ are mass and reduced Planck constant respectively. Applying WKB approximation, the vector field $ \Psi_\mu $ takes the form as
\begin{eqnarray}
\Psi_\mu=C_\mu {{\rm \exp}[\frac{i}{\hbar}S(t, r, \theta, \phi})].
\end{eqnarray}
Using Eq. (10) in Eq. (9) and neglecting higher powers of $ \hbar $, we obtain as
\begin{eqnarray}
g^{\mu \lambda}g^{\nu \rho}(-C_\rho\partial_\mu S\partial_\lambda S+C_\lambda\partial_\mu S\partial_\rho S)+m^2g^{\nu \lambda}C_\lambda=0,
\end{eqnarray}
where $\rho, \mu, \lambda = 0,1,2,3$.  Substituting  Eq. (5) in Eq. (9), the following four equations are obtained as
\begin{eqnarray}
&&g^{tt}\Big\{-g^{rr}\Big(\frac{\partial S}{\partial r}\Big)^2-g^{\theta \theta}\Big(\frac{\partial S}{\partial \theta}\Big)^2-g^{\phi \phi}\Big(\frac{\partial S}{\partial \phi }\Big)^2+m^2\Big\}C_0+g^{rr}g^{tt}\frac{\partial S}{\partial t} \frac{\partial S}{\partial r}C_1\cr &&+g^{\theta \theta}g^{tt}\frac{\partial  S}{\partial t} \frac{\partial S}{\partial \theta} C_2+g^{tt}g^{\phi \phi}\frac{\partial S}{\partial t} \frac{\partial S}{\partial \phi}  C_3=0,
\end{eqnarray}
\begin{eqnarray}
&&g^{rr}g^{tt}\frac{\partial S}{\partial t} \frac{\partial S}{\partial r} C_0+g^{rr}\Big\{-g^{tt}\Big(\frac{\partial S}{\partial t}\Big)^2-g^{\theta \theta}\Big(\frac{\partial S}{\partial \theta} \Big)^2-g^{\phi \phi}\Big(\frac{\partial S}{\partial \theta}\Big)^2+m^2\Big\}C_1\cr &&+g^{rr}g^{\theta \theta}\frac{\partial S}{\partial r} \frac{\partial S}{\partial \theta } C_2 + g^{rr}g^{\phi\phi}\frac{\partial S}{\partial r}\frac{\partial S}{\partial \theta}  C_3=0,
\end{eqnarray}
\begin{eqnarray}
&&g^{\theta \theta}g^{tt}\frac{\partial S}{\partial t}\frac{\partial S}{\partial\theta} C_0 +g^{\theta\theta}g^{rr}\frac{\partial S}{\partial r} \frac{\partial S}{\partial \theta} C_1+g^{\theta\theta}\Big\{-g^{tt}\Big(\frac{\partial S}{\partial t}\Big)^2-g^{rr}\Big(\frac{\partial  S}{\partial r}\Big)^2-\cr &&g^{\phi \phi}\Big(\frac{\partial S}{\partial \phi}\Big)^2 +m^2\Big\}C_2+g^{\theta\theta}\frac{\partial S}{\partial \theta }\frac{\partial S}{\partial \phi } C_3=0,
\end{eqnarray}

\begin{eqnarray}
&&g^{tt}g^{\phi\phi}\frac{\partial S}{\partial t}\frac{\partial S}{\partial \phi}C_0+g^{rr}g^{\phi \phi}\frac{\partial S}{\partial r}\frac{\partial S}{\partial \phi }C_1 + g^{\theta \theta}g^{\phi \phi}\frac{\partial S}{\partial \theta }\frac{\partial S}{\partial \phi} C_2+g^{\phi\phi}\Big\{-g^{tt}\Big(\frac{\partial S}{\partial r}\Big)^2\cr&& -g^{rr}\Big(\frac{\partial S}{\partial r}\Big)^2-g^{\theta\theta}\Big(\frac{\partial S}{\partial \theta}\Big)^2+m^2\Big\}C_3=0.
\end{eqnarray}

The above four equations contain the four variables namely $ t $, $ r $, $ \theta $ and $ \phi $. We know that the Hawking radiation  takes place along the radial direction only. To obtain radial equation for studying of Hawking radiation near the event horizon of black hole, the action $ S $ can be written as 
\begin{equation}
S=-\omega t+W(r)+j\phi+\Theta(\theta) + \zeta,
\end{equation}
where $\omega$ and $j$ denote the energy and angular momentum of the emitted vector particle respectively and $ \zeta $ is the complex constant. Then we have,
\begin{eqnarray}
&&g^{tt}(-g^{rr}W'^2-g^{\theta\theta}{\Theta '}^2-g^{\phi\phi}j^2+m^2)C_0-g^{rr}g^{tt}\omega W'C_1\cr && -g^{\theta\theta}g^{tt}\omega \Theta' C_2-g^{tt}g^{\phi\phi}\omega jC_3=0,\\
&&-g^{tt}g^{rr}\omega W'C_0+g^{rr}(-g^{tt}\omega^2-g^{\theta\theta}{\Theta'}^2-g^{\phi\phi}j^2+m^2)C_1\cr && +g^{rr}g^{\theta\theta}W'\Theta'C_2+g^{rr}g^{\phi\phi}W'\Theta'C_3=0,\\
&& -g^{tt}g^{\theta\theta}\omega\Theta'C_0+g^{rr}g^{\theta\theta}W'\Theta' C_1+g^{\theta\theta}(-g^{tt}\omega^2-g^{rr}W'^2-g^{\phi\phi}j^2\cr && +m^2)C_2+g^{\theta\theta}g^{\phi\phi}\Theta' j C_3=0,\\
&& -g^{tt}g^{\phi\phi}\omega jC_0+g^{rr}g^{\phi\phi}W'jC_1+g^{\theta\theta}g^{\phi\phi}\Theta' jC_2+g^{\phi\phi}(-g^{tt}\omega^2\cr&& -g^{rr}W'^2-g^{\theta\theta}{\Theta'}^2+m^2)C_3=0,
\end{eqnarray}
where $\Theta'=\frac{\partial \Theta}{\partial \theta}$ and  $W'=\frac{\partial W}{\partial r}$. The above four equations can be treated as a matrix equation $K(C_0,C_1,C_2,C_3)^T=0$. The Eqs. (17-20) will have non-trivial solution iff the determinant of the coefficient matrix $ K $ is zero. The value of the determinant of $ K $ is 
\begin{eqnarray}
{\rm det} K =\frac{[z\sin^2\theta\{\Delta\Delta_\theta(\Delta\Delta_\theta W'^2-\rho^2m^2+\Delta_\theta \Theta'^2)-z\Xi^2 \omega^2\}+\Delta\Delta_\theta j^2\rho^4\Xi^2]^3}{m^{-2}\Xi^{-2}z^3\Delta^3\Delta_\theta^3\rho^{10}\sin^8\theta},
\end{eqnarray}
where $z=\Delta_\theta(r^2+a^2)^2-\Delta a^2 \sin^2\theta$. Putting det$ K $ is equal to zero, we get
\begin{eqnarray}
W'=\pm\sqrt{\frac{\frac{Y\Xi^2\omega^2}{\Delta_\theta}-\Delta( \Delta_\theta \theta'^2-\rho^2 m^2+\frac{j^2\rho^4\Xi^2}{Y\sin^2\theta})}{\Delta}}.
\end{eqnarray}
For KNdS black hole, there is a singularity around the event horizon $ r=r_{h} $. Integrating the above equation and applying residue theorem of complex analysis at the pole $ r=r_{h} $, we get 
\begin{eqnarray}
{\rm Im}W_{\pm}=\pm\frac{\pi\Xi \omega(r_h^2+a^2)}{2r_h-\frac{4}{3}\Lambda r_h^3-\frac{2}{3}\Lambda r_h a^2-2M},
\end{eqnarray}
where $ W_{+} $ and $ W_{-} $ indicate the radial action of the outgoing particle and the ingoing particle respectively. It is known that the imaginary part of the action is related to the tunneling probabilities of particles along the classically forbidden trajectory in accordance with WKB approximation. The probability of the particle crossing the event horizon $ r=r_{h} $ of black hole are defined by
\begin{eqnarray}
P_{\rm out} &=& \exp[-\frac{2}{\hbar}{\rm Im} S] = \exp[-\frac{2}{\hbar}({\rm Im} W_{+} + {\rm Im} \zeta)]
\end{eqnarray}
and 
\begin{eqnarray}
P_{\rm in} &=& \exp[-\frac{2}{\hbar}{\rm Im} S] = \exp[-\frac{2}{\hbar}({\rm Im} W_{-} + {\rm Im} \zeta)].
\end{eqnarray}
The ingoing spin-1 particle has a $ 100\% $ chance to enter the black hole in accordance with semiclassical approximation. It indicates that $  {\rm Im} \zeta =- {\rm Im} W_{-}$. Since $ W_{-}=W_{+} $, the probability of outgoing particles is computed as
\begin{eqnarray}
\Gamma_{\rm rate}=\frac{\Gamma_{\rm emission}}{\Gamma_{\rm absorption}}&=&  \exp(-4{\rm Im}W_+)\cr &=& \exp\Bigg[-\frac{2\Xi \omega \pi(r_h^2+a^2)}{r_h-M-\frac{2}{3}\Lambda r_h^3-\frac{\Lambda}{3}r_h a^2}\Bigg].
\end{eqnarray}
From Eq. (26), the emission spectrum of massive vector particle near the event horizon $ r=r_{h} $ of KNdS black hole is derived as\cite{damourr}
\begin{eqnarray}
N(\omega) = \frac{1}{e^{\frac{\omega}{T_{H}}-1}},
\end{eqnarray}
with the standard Hawking temperature of KNdS black hole at the event horizon by 
\begin{eqnarray}
T_H=\frac{1}{2\pi}\Bigg[\frac{r_h-M-\frac{2}{3}\Lambda r_h^3-\frac{\Lambda}{3}r_h a^2}{ \Xi (r_h^2+a^2)}\Bigg],
\end{eqnarray}
which is consistent with the actual calculation given in Eq. (8).
 
\section{Review of the deformed Hamilton-Jacobi equation}
To investigate the string theory and quantum gravity, the deformed dispersion relation in the magnitude of Planck scale is given [39-45] by
\begin{eqnarray}
p_{0}^{2} = \overrightarrow{p}^{2} + m^{2} - (Lp_{0})^{\alpha}\overrightarrow{p}^{2},
\end{eqnarray}
where $ p_{0} $ and $ \overrightarrow{p} $ are the energy and momentum of the particle respectively. $ m $ and $ L $ are the static mass and arbitrary constant in the magnitude of the Planck scale. The constant term $ L $ comes from the Lorentz invariance violation theory. In the Liouville string theory mode, the value of $ \alpha $ is taken as unity. Kruglov \cite{sikpl} derived the modified form of Dirac equation from Eq. (29) for $ \alpha = 2$. In flat space time, the Dirac equation with spin-1/2 fermions can be written as
\begin{eqnarray}
\Big(\overline{\gamma}^{\mu}\partial_{\mu} + \frac{m}{\hbar} - iL\overline{\gamma}^{t}\partial_{t}\overline{\gamma}^{j}\partial_{j}\Big)\psi=0.
\end{eqnarray}
 Rarita and Schwinger [47] derived the general fermion equation in flat spacetime as
\begin{eqnarray}
\Big(\overline{\gamma}^{\mu}\partial_{\mu} + \frac{m}{\hbar}\Big)\psi_{\alpha_{1}\dots\alpha_{k}} = 0,
\end{eqnarray}
which obeys the conditions $ \overline{\gamma}^{\mu}\psi_{\mu\alpha_{2}\dots\alpha_{k}}= \partial_{\mu}\psi_{\alpha_{2}\dots\alpha_{k}}^{\mu} = \psi_{\mu\alpha_{3}\dots\alpha_{k}}^{\mu} =0.$ In curved spacetime, the Rarita-Schwinger equation can be expressed as
\begin{eqnarray}
\Big(\gamma^{\mu}D_{\mu} + \frac{m}{\hbar}\Big)\psi_{\alpha_{1}\dots\alpha_{k}} = 0,
\end{eqnarray}
which satisfies the conditions $ \gamma^{\mu}\psi_{\mu\alpha_{2}\dots\alpha_{k}}= D_{\mu}\psi_{\alpha_{2}\dots\alpha_{k}}^{\mu} = \psi_{\mu\alpha_{3}\dots\alpha_{k}}^{\mu} =0.$  Putting $ \alpha =2 $ in Eq. (29), we get the Rarita-Schwinger equation in flat spacetime as
\begin{eqnarray}
\Big(\overline{\gamma}^{\mu}\partial_{\mu} + \frac{m}{\hbar} - \sigma \hbar \overline{\gamma}^{t}\partial_{t}\overline{\gamma}^{j}\partial_{j}\Big)\psi_{\alpha_{1}\dots\alpha_{k}}=0.
\end{eqnarray} 
For the small correction on quantum scale, we take   $ \sigma\ll 1 $, so the term $\sigma \hbar \gamma^{t}D_{t}\gamma^{j}D_{j}  $ is an infinitesimal quantity. Then the Rarita-Schwinger equation in curved space becomes 
\begin{eqnarray}
\Big(\gamma^{\mu}D_{\mu} + \frac{m}{\hbar} - \sigma \hbar \gamma^{t}D_{t}\gamma^{j}D_{j}\Big)\psi_{\alpha_{1}\dots\alpha_{k}}=0, 
\end{eqnarray}
where $ D_{\mu} = \partial_{\mu} + \Omega_{\mu} +(i/\hbar)eA_{\mu} $ is called the operator of covariant derivative in gravitational background, $ \Omega_{\mu} $ is the spin connection in curve spacetime and $ \gamma^{\mu} $ satisfies the anticommutation relation which is defined as
\begin{eqnarray}
\{\gamma^{\mu}, \gamma^{\nu}\} = 2g^{\mu \nu}I,
\end{eqnarray}
where $ I $ is a identity matrix. The Eq. (33) will become the Dirac equation of spin-1/2 for $ k=0 $. For finding the fermions tunneling radiation near the event horizon of black hole, the wave function is taken as  
\begin{eqnarray}
\psi_{\alpha_{1}\dots\alpha_{k}} = \xi_{\alpha_{1}\dots\alpha_{k}}e^{\frac{i}{\hbar}S},
\end{eqnarray}
where $ S $ is called the action of fermions and it can be defined as 
\begin{eqnarray}
S= -\omega t + S_{0}(\bar{r}, \theta, \phi),
\end{eqnarray}
where $ \partial_{\psi}S = j$ and $ \partial_{t}S = -\omega$. $ j $ and $ \omega $ are the angular momentum and radiant energy of the emitting particles. Using Eq. (36) in Eq. (34),    neglecting the higher order terms of $ \hbar $, we  obtain 
\begin{eqnarray}
i\gamma^{\mu}(\partial_{\mu}S + eA_{\mu})\xi_{\alpha_{1}\dots\alpha_{k}} + m\xi_{\alpha_{1}\dots\alpha_{k}} - \sigma \gamma^{t}(\omega -eA_{t})\gamma^{j}(\partial_{j}S +eA_{j})\xi_{\alpha_{1}\dots\alpha_{k}} =0.
\end{eqnarray} 
By considering the condition $ (\partial_{\mu}S + eA_{\mu}) = -\gamma^{t}(\omega -eA_{t})  + \gamma^{j}(\partial_{\j}S + eA_{\j})$ in the Eq. (38),
the deformed Hamilton-Jacobi equation is obtained as [48, 49] 
\begin{eqnarray}
g^{ab}(\partial_{a}S+eA_{a})(\partial S+eA_{b})+m^{2}-2\sigma m g^{tt}(\omega-eA_{t})^{2}=0.
\end{eqnarray}
The above equation is also known as the modified form of Rarita-Schwinger equation. It has been shown that the deformed Hamilton-Jacobi equation of boson is different from the result of fermions [50]. Therefore Eq. (39) can be taken as the deformed Hamilton-Jacobi equation of the fermions.
\section{Modified Hawking temperature for stationary KNdS black hole}

According to the no-hair law of black hole, the (3+1) dimensional KNdS black hole exhibits the four characteristics, namely mass of the black hole, charge, angular momentum and cosmological constant. In KNdS black hole, the three conditions of infinite spacetime will exist, namely asymptotic flat space, asymptotic de Sitter space and asymptotic anti de Sitter space. To investigate the modified Hawking temperature near the event horizon of static and stationary KNdS black hole,  we substitute Eq. (1) in Eq. (39) and the following equation is obtained as 
\begin{eqnarray}
&&- \Xi^2\Big[\frac{(r^2+a^2)^2}{\Delta \rho^2}-\frac{a^2\sin^2\theta}{\Delta_\theta \rho^2}\Big]\omega^2+\frac{\Delta}{\rho^2}\partial_rS^2+\frac{\Delta_\theta}{\rho^2}\partial_\theta S^2+\frac{\Xi^2}{\rho^2}\Big[\frac{1}{\Delta_\theta \sin^2\theta}-\frac{a^2}{\Delta}\Big]j^2 \cr&& +2\frac{a\Xi^2}{\rho^2}\Big[\frac{(r^2+a^2)}{\Delta}-\frac{1}{\Delta_\theta}\Big]j\omega + m^2+2 \sigma m\omega^2\frac{\Xi^2}{\rho^2}\Big[\frac{(r^2+a^2)^2}{\Delta}-\frac{a^2\sin^2\theta}{\Delta_\theta}\Big]=0. 
\end{eqnarray}
 Using the separation of variables, the radial and angular equations are derived from Eq. (40), but the angular equation does not give any physical meaning. The radial equation is given by
\begin{eqnarray}
\frac{\Xi^2}{\Delta}\Big[B\omega-aj\Big]^2-\Delta \partial_rS^2-\frac{2\sigma m \omega^2 \Xi^2 B^2}{\Delta}=0 .
\end{eqnarray}
The required solution of the above equation is
\begin{eqnarray}
 \partial_rS=\pm\frac{\Xi}{f(r)}(\omega-j\Omega_0)\sqrt{1-2\sigma m \Theta^2}.
\end{eqnarray}
where 
\begin{eqnarray}
f(r)=\frac{\Delta}{r^2+a^2},\,\,\, \Omega_0=\frac{a}{r^2+a^2},\,\,\,\Theta=\frac{\omega(r^2+a^2)}{\omega(r^2+a^2)-ja}.
\end{eqnarray}
On integration, the radial action can be calculated from Eq. (42) as
\begin{eqnarray}
S_{\pm }(r)=\pm \int \limits_{0}^{\infty} \frac{\Xi}{f(r)}(\omega-j\Omega_0)\sqrt{1-2\sigma m \Theta^2}dr.
\end{eqnarray}
The KNdS black hole has four singularities [36]. We will consider the modified Hawking temperature at the event horizon $r=r_h$ only of KNdS black hole. We choose the limits of integration such that the particle moves through the exterior of event horizon. Applying Feynman prescription around the pole $r=r_h$ and completing the integral, the imaginary part of the action is calculated as
\begin{eqnarray}
{\rm Im}S_{\pm} = \pm \frac{\pi\Xi (r_{h}^2+a^2)}{2\Delta^{\prime}r_h}(\omega-j\Omega_0)(1-\sigma m \Theta^2),
\end{eqnarray}
where $S_{+}$ corresponds to the outgoing particle (moving away from the black hole) and $S_{-}$ corresponds to the ingoing particle (moving toward the black hole). The probabilities that the particle can cross the event horizon of black hole are

\begin{eqnarray}
P_{\rm out}=\exp[\frac{2}{\hbar}{\rm Im} S_{+}]
\end{eqnarray}
and
\begin{eqnarray}
P_{\rm in}=\exp[\frac{2}{\hbar}{\rm Im} S_{-}].
\end{eqnarray}
There is a $100 \%$ chance for the ingoing particle to enter the black hole according to the semiclassical approximation. This means that $S_+=-S_-$. The probability of outgoing particle at the event horizon of KNdS black hole is derived as
\begin{eqnarray}
\Gamma &=& \exp(-2({\rm Im} S_{+}-{\rm Im} S_{-}))\cr
&=&\exp\{-2\frac{\pi \Xi (r_{h}^2+a^2)}{\Delta^{\prime}(r_h)}(\omega-j\Omega_0)(1-\sigma m \Theta^2)\}.
\end{eqnarray}
The modified Hawking temperature at the event horizon $r=r_h$ is given by
\begin{eqnarray}
T&=&\frac{\Delta^{\prime}r_h}{2\pi(1-\sigma m \Theta^2)(r_{h}^2+a^2)\Xi}\cr
&=& T_H(1+\sigma m \Theta^2),
\end{eqnarray}
where the original Hawking temperature is
\begin{eqnarray}
T_{H}=\frac{1}{2\pi}\Bigg[\frac{r_h-M-\frac{2}{3}\Lambda r_h^3-\frac{\Lambda}{3}r_h a^2}{ \Xi (r_h^2+a^2)}\Bigg].
\end{eqnarray}
From Eq. (49), the Hawking temperature near the event horizon of stationary KNdS black hole is modified due to presence of correction term $\sigma m \Theta^2$. Since $ m>0$, $\sigma>0$ and $\Theta>0$, Eq. (49) shows that the modified Hawking temperature near the event horizon of KNdS black hole is greater than the original Hawking temperature given in Eq. (50). The modified Hawking temperature depends not only on $m\sigma\Theta^2$ but also on black hole mass, cosmological constant and on the rotational parameter. The value of $\Theta$ depends on the angular momentum parameter and the radiant energy parameter. For nonrotating black hole, the value of $\Theta$ tends to unity. It is worth mentioning that near the horizon, $ \Theta $ is independent of angular coordinate 
$ \theta $. $\Theta$ determines the position of horizon and angular coordinate $ \theta $. From Eqs. (8), (28) and (50), the Hawking temperatures near the event horizon of stationary KNdS black hole are equal.
\section{ Modified Hawking radiation for nonstationary KNdS black hole}
In the retarded time coordinate, the metric of nonstationary KNdS black hole is defined as
\begin{eqnarray}
ds^2=&\frac{1}{\rho^2\Xi^2}[\Delta_\lambda-\Delta_\theta a^2\sin^2\theta]du^2+\frac{2}{\Xi}[du-a\sin^2\theta d\phi]dr\cr & -\frac{\rho^{2}}{\Delta_\theta}d\theta^2+\frac{2a}{\rho^{2}\Xi^{2}}[\Delta_\theta(r^2+a^2)-\Delta_\lambda]\sin^2\theta dud\phi \cr & -\frac{1}{\rho^{2}\Xi^{2}}[\Delta_\theta(r^2+a^2)^2-\Delta_\lambda a^2\sin^2\theta]\sin^2\theta d\phi^2,
\end{eqnarray}
where $\rho^2$, $\Xi$ and $\Delta_\theta$ are given in Eq. (2). The term $\Delta_\lambda$ is given by
\begin{eqnarray}
 \Delta_\lambda = r^2+a^2-2M(u)r+Q^2(u)- \frac{1}{3}\Lambda r^2(r^2+a^2).
\end{eqnarray}
$M(u)$  and $Q(u)$ are the mass and charge of the nonstationary KNdS black hole, respectively, as seen by the observer at infinity, and they are arbitrary functions of the retarded time coordinate $u$. The four vector electromagnetic potential $ A_{\mu} $ of KNdS black hole is given by
\begin{eqnarray}
A_{\mu} =\Big[\frac{Qr\sqrt{\Xi}}{\rho^2},0,0,-\frac{Qr\sqrt{\Xi }a\sin^{2}\theta}{\rho^{2}}\Big].
\end{eqnarray}
 The event horizon of nonstationary KNdS black hole is characterized by null hypersurface condition: $ f(u,r,\theta,\phi) =0 $. The position of horizon of stationary or nonstationary black hole is obtainable from null hypersurface condition. 
 The general expression of null hypersurface equation is
\begin{eqnarray}
&g^{\mu \nu}\frac{\partial f}{\partial x^\mu}\frac{\partial f}{\partial x^\nu}=0,
\end{eqnarray} 
We know that the event horizon of nonstationary KNdS black hole is changed with retarded time coordinate $u=t-r_*$ and different angles $\theta$, $\phi$.  The tortoise coordinate describes the space time geometry outside the event horizon of nonstationary KNdS black hole and in this case, $r_*$ will be positive infinity when tending to infinite point and $r_*$ approaches to negative infinity near the event horizon. The space time geometry of nonstationary KNdS black hole is symmetric about $\phi$ axis. To investigate the tunneling rate of fermions near the event horizon of black hole, the general tortoise coordinate transformation is given by [51-57]
\begin{eqnarray}
&&r_* = r+ \frac{1}{2\kappa(u_0, \theta_0,\phi_{0})}ln\Bigg(\frac {r-r_h(u, \theta\phi)}{r_h(u, \theta,\phi)}\Bigg)\cr 
&& u_*=u-u_0, 
     \,\,\,\,\,\,\,\,\,\, \theta_*= \theta-\theta_0, \,\,\,\,\,\,\,\,\,    \phi_*=\phi-\phi_0,
\end{eqnarray}
where $u_0$, $\theta_0$, $\phi_0$ are the parameters under the tortoise coordinate transformation. From Eq. (55), we  get
\begin{eqnarray}
&&\frac{\partial}{\partial r}=\Bigg\{1+\frac{1}{2\kappa (r-r_h)}\Bigg\}\frac{\partial}{\partial r_*}\cr 
&&\frac{\partial}{\partial u}=\Bigg\{ \frac{\partial}{\partial u_*}-\frac{rr_{h,u}}{2\kappa r_h(r-r_h)}\frac{\partial}{\partial r_*}\Bigg\}\cr 
&& \frac{\partial}{\partial \theta}=\Bigg\{\frac{\partial}{\partial \theta_*}-\frac{rr_{h,\theta}}{2\kappa r_h(r-r_h)}\frac{\partial}{\partial r_*}\Bigg \}\cr
&&\frac{\partial}{\partial \phi}=\Bigg\{\frac{\partial}{\partial \phi_*}-\frac{rr_{h,\phi}}{2\kappa r_h(r-r_h)}\frac{\partial}{\partial r_*}\Bigg\},
\end{eqnarray}
where $r_{h,u}=\frac{\partial r_h}{\partial u}$, $r_{h,\theta}=\frac{\partial r_h}{\partial \theta}$ and $r_{h,u}=\frac{\partial r_h}{\partial \phi } $.
  $ r_{h,u} $ indicates the rate of evaporation near the horizon of KNdS black hole. The expansion of event horizon is occurred for $ \partial r_{h}/ \partial u > 0$ (absorbing black hole) and the contraction of event horizon is occurred  for $ \partial r_{h}/ \partial u < 0$. The terms $ r_{h, \theta} =\partial r_{h}/\partial \theta $ and $ r_{h, \phi} =\partial r_{h}/\partial \phi $ represent the rate of event horizon varying with angles. They indicate the rotation effect of nonstationary KNdS black hole. $ r_{h} $  represents the location of event horizon and relies on retarded time $ u_{0} $ and angular coordinates $ \theta_{0} $ and $ \phi_{0} $. $ \kappa \equiv\kappa(u_{0}, \theta_{0}, \phi_{0}) $ is the surface gravity of black hole which is dependent on retarded time and angular coordinates. From Eqs. (51), (54) and (55) and taking limit $ u \rightarrow u_{0}, r\rightarrow r_{h}, \theta\rightarrow\theta_{0} $ and $ \phi\rightarrow\phi_{0} $, where $u_0$, $\theta_0$ and $\phi_0$ are initial state of the hole, then the event horizon equation of nonstationary KNdS black hole is calculated as
\begin{eqnarray}
&&\frac{a^2 \Xi^2\sin^2 \theta r_{h,u}^2}{\Delta_\theta}+2(r_h^2+a^2)\Xi r_{h,u}
+\frac{2a\Xi^2r_{h,u}r_{h,\phi}}{\Delta_\theta}+2a\Xi r_{h,u}+\Delta_\lambda(r_{h})\cr &&+\Delta_\theta r_{h,\theta}^2+\frac{\Xi^2 r_{h,\phi}^2}{\Delta_\theta \sin^2\theta}=0,
\end{eqnarray}
where $ \Delta_\lambda(r_{h}) = r_{h}^2+a^2-2M(u)r_{h}+Q^2(u)- \frac{1}{3}\Lambda r_{h}^2(r_{h}^2+a^2)  $. Eq. (57) indicates that the shapes of horizons are dependent on retarded time $u_0$ and angular coordinates $\theta_0$, $\phi_0$.  From Eqs. (39) and (51), we have the equation of motion of the half-integer fermions in nonstationary KNdS black hole as
\begin{eqnarray}
&g^{00}\Big(\frac{\partial S}{\partial u}+eA_0\Big)^2+2g^{01}\Big(\frac{\partial S}{\partial u}+eA_0\Big)\Big(\frac{\partial S}{\partial r}\Big)+2g^{03}\Big(\frac{\partial S}{\partial u}+eA_0\Big)\Big(\frac{\partial S}{\partial \phi}+eA_3\Big)\cr& +2g^{13}\Big(\frac{\partial S}{\partial r}\Big)\Big(\frac{\partial S}{\partial \phi}+eA_3\Big)+g^{11}\Big(\frac{\partial S}{\partial r}\Big)^2+g^{22}\Big(\frac{\partial S}{\partial\theta}\Big)^2+g^{33}\Big(\frac{\partial S}{\partial \phi}+eA_3)^2+m^2\cr&
 -2\sigma mg^{00}\Big(\frac{\partial S}{\partial u}-eA_0\Big)^2=0.
\end{eqnarray}
Eq. (58) is the equation of motion of fermions in nonstationary KNdS black hole where $S=S(u, r, \theta,\phi)$. From Eqs. (56) and (58), the equation  of fermions in nonstationary KNdS black hole can be written as
\begin{eqnarray}
\frac{A}{D}\Bigg(\frac{\partial S}{\partial r_*}\Bigg)^2+2\frac{\partial S}{\partial u_*}\frac{\partial S}{\partial r_*}+2\frac{B}{D}\frac{\partial S}{\partial r_*}+2\kappa(r-r_h)\frac{C}{D}=0,
\end{eqnarray}
where the terms $A$, $B$, $C$ and $D$ are defined by
\begin{eqnarray}
A&=&\frac{1}{2\kappa(r-r_h)}\Big[\frac{g^{00}r^2r_{h,u}^2}{r_h^2}-\frac{2g^{01}rr_{h,u}\{{2\kappa(r-r_h)+1}\}}{r_h}+\frac{2g^{03}r^2r_{h,u}r_{h,\phi}}{r_h^2}\cr&& -\frac{2g^{13}\{2\kappa(r-r_h)+1\}r r_{h,\phi}}{r_h}+g^{11}\{2\kappa (r-r_h)+1\}^2+\frac{g^{22}r^2r_{h,\theta}^2}{r_h^2}+\frac{g^{33}r^2r_{h,\phi}^2}{r_h^2}\Big]\cr
B&=&-\frac{eA_0g^{00}rr_{h,u}}{r_h}+eA_0g^{01}\{2\kappa(r-r_h)+1\}-\frac{g^{03}rr_{h,u}p_\phi}{r_h}-\frac{eA_3g^{03}rr_{h,u}}{r_h}\cr && -\frac{eA_0g^{03}rr_{h,\phi}}{r_h}+g^{13}\{2\kappa (r-r_h)+1\}p_\phi+eA_3g^{13}\{2\kappa(r-r_h)+1\}\cr &&-\frac{g^{22}rr_{h,\theta} p_\theta}{r_h} -\frac{g^{33}rr_{h,\phi}p_\phi}{r_h}-\frac{eA_3g^{33}rr_{h,\phi}}{r-r_{h}}-\frac{2\sigma meA_0g^{00}rr_{h,u}}{r_h}\cr
 C&=&g^{00}(\omega-eA_0)^2 - 2g^{03}(\omega-eA_0)(p_\phi+eA_3)+g^{22}p_\theta^2+g^{33}(p_\phi+eA_3)^2\cr &&+m^2-2\sigma mg^{00}(\omega+eA_0)^2\cr 
D&=&-\frac{g^{00}rr_{h,u}}{r_h}+g^{01}\{2\kappa(r-r_h)+1\}-\frac{g^{03}rr_{h,\phi}}{r_h}+\frac{2rr_{h,u}\sigma mg^{00}}{r_h}.
\end{eqnarray}
 For KNdS black hole in retarded time coordinate, Eq. (37) can be written as  $ S = -\omega u_{*} + S_{0}(r_{*}, \theta_{*},\phi_{*}) $ and we defined as
\begin{eqnarray}
 \frac{\partial S}{\partial u_*}=-\omega, \,\,\,\,\,\,\,
\frac{\partial S}{\partial\theta_*}=p_\theta, \,\,\,\,\,\, \frac{\partial S}{\partial \phi_*}=p_\phi,
\end{eqnarray}
where $\omega$, $p_{\theta}$ and $p_{\phi}$ denote the energy of fermion tunneling radiation, $\theta$ component of the generalized momentum of fermion and  $\phi$ component of the generalized momentum of fermion respectively. 
To investigate the modified surface gravity and modified Hawking temperature near the event horizon of nonstationary black hole, it is assumed that the coefficient of $ \big(\partial S/\partial r_*\big)^2 $ tends to unity when $ r \rightarrow r_{h}, u \rightarrow u_{0}, \theta \rightarrow \theta_{0} $ and $ \phi \rightarrow \phi_{0} $. Here we get an infinite limit of $0/0$ type near the event horizon. Applying the L'Hopital's rule, then the modified surface gravity near the event horizon is given by
\begin{eqnarray}
\kappa=\frac{r_h\big(1+2\Xi r_{h,u}\big)-M-\frac{2}{3}\Lambda r_h^3-\frac{\Lambda}{3} r_ha^2-r_h^{-1}\big\{\Delta_\lambda+r_{h,u}\Xi(r_h^2+a^2)+a\Xi r_{h,\phi}\big\}}{\Xi(r_h^2+a^2)(1-2r_{h,u})+r_{h,\phi}\big(\frac{a\Xi^{2}}{\Delta_\theta}-2\Xi a\big)-2\Delta_\lambda},\nonumber\\
\end{eqnarray}
where $L=a^2\Delta_\theta^{-1}\Xi^2\sin^2\theta_0r_{h,u}(1-2\sigma m)$.
 Using Eq. (57) to the above equation, the modified surface gravity can be expressed as 
 \begin{eqnarray}
\kappa =\frac{r_h\big(1+2\Xi r_{h,u}\big)-M-\frac{2}{3}\Lambda r_h^3-\frac{\Lambda}{3} r_h a^2-r^{-1}_h\Big\{\Delta_\lambda+r_{h,u} \Xi(r_h^2+a^2)+ a\Xi r_{h,\phi}\Big\} }{\Big\{\Xi (r_h^2+a^2)+\frac{a^2\Xi^{2}\sin^2\theta_0r_{h,u}}{\Delta_\theta}\Big\}(1+2r_{h,u})+Z},\nonumber\\
\end{eqnarray}
 where $Z=2\Delta_\theta r_{h,\theta}^2+r_{h,\phi}\Big(\frac{4a\Xi^2 r_{h,u}}{\Delta_\theta}+\frac{2\Xi^{2}r_{h,\phi}}{\Delta_\theta \sin^2\theta_{0}}+2a\Xi+\frac{a\Xi^{2}}{\Delta_\theta}\Big)-\frac{2\sigma ma^2\Xi^{2}\sin^2\theta_{0} r_{h,u}}{\Delta_\theta}.$\\ \\ \\ 
The surface gravity near the event horizon of nonstationary KNdS black hole is modified due to presence of $2\Delta^{-1}_\theta\sigma ma^2\Xi^{2}\sin^2\theta_{0} r_{h,u}$ in Eq. (63) and the modified surface gravity depends on the properties of the event horizon, the mass of the black hole $ M $, cosmological constant $ \Lambda $, charge $ Q $, retarded time $ u_{0} $, angular coordinates $\theta_{0}$, $\phi_{0}$. Near the event horizon, Eq. (59) can be written as
\begin{eqnarray}
\Bigg(\frac{\partial S}{\partial r_*}\Bigg)^2+2(\omega -\omega_0)\frac{\partial S}{\partial r_*}=0,
\end{eqnarray}
where $\frac{B}{D}=\omega_0$ when  $ r \rightarrow r_{h}, u \rightarrow u_{0}, \theta \rightarrow \theta_{0} $ and $ \phi \rightarrow \phi_{0} $. The value of $\omega_0$ is
\begin{eqnarray}
\omega_0&=&\Big[-eA_0g^{00}r_{h,u}+eA_0g^{01}-(p_\phi+eA_3)g^{03}r_{h,u}-eA_0g^{03}r_{h,\phi}\cr && +(p_\phi+eA_3)g^{13}-g^{22}r_{h,\theta}p_\theta-(p_\phi+eA_3)g^{33}-2\sigma mg^{00}eA_0r_{h,u}\Big]\cr && \times\frac{1}{(g^{01}-g^{00}r_{h,u}-g^{03}r_{h,\phi}+2\sigma mg^{00}r_{h,u})}.
\end{eqnarray}
  The chemical potential of nonstationary KNdS black hole is modified due to presence of two terms $2\sigma mg^{00}eA_0r_{h,u}$ and $2\sigma mg^{00}r_{h,u}$ in Eq. (65). 
  $\omega_0$ is the exact corrected chemical potential near the event horizon of nonstationary KNdS black hole. The modified chemical potential depends not only on Mass of the black hole, but also on cosmological constant $\Lambda$, charge $Q$, generalized momentum of the fermions, retarded time and on different angles.  From Eq. (56), the first equation can be written as 
 
 \begin{eqnarray}
\frac{\partial S}{\partial r_*}=\frac{[2\kappa(r-r_h)+1](\omega-\omega_0)\pm(\omega-\omega_0)}{[2\kappa(r-r_h)]}.
\end{eqnarray}
The above equation has a singularity around the event horizon  $r=r_h$ of black hole. On integration of Eq. (65) above the semicircle $r=r_h$ by applying Feynman prescription, the imaginary part of the action is calculated as
\begin{eqnarray}
{\rm Im S}_{\pm}=\frac{\pi}{2\kappa}[(\omega-\omega_{0})\pm(\omega-\omega_{0})],
\end{eqnarray}
where $S_{+}$ and $S_{-}$ are the outgoing fermions and ingoing fermions near the horizon of KNdS black hole. Considering both outgoing and ingoing fermions, the total tunneling probability across the event horizon of KNdS black hole is 
\begin{eqnarray}
\Gamma&=&\frac{\Gamma_{\rm emission}}{\Gamma_{\rm absorption}}=\exp \Big[-\frac{(\omega-\omega_{0})}{T_h}\Big],
\end{eqnarray}
where the modified Hawking temperature at the event horizon of black hole is 
\begin{eqnarray}
T_h =&&\frac{1}{2\pi}\Bigg[\frac{r_h\big(1+2\Xi r_{h,u}\big)-M-\frac{2}{3}\Lambda r_h^3-\frac{\Lambda}{3} r_h a^2}{\Big\{\Xi (r_h^2+a^2)+\frac{a^2\Xi^{2}\sin^2\theta_0r_{h,u}}{\Delta_\theta}\Big\}(1+2r_{h,u})+Z}\cr&&-\frac{r^{-1}_h\Big\{\Delta_\lambda+r_{h,u} \Xi(r_h^2+a^2)+ a\Xi r_{h,\phi}\Big\}}{\Big\{\Xi (r_h^2+a^2)+\frac{a^2\Xi^{2}\sin^2\theta_0r_{h,u}}{\Delta_\theta}\Big\}(1+2r_{h,u})+Z}\Bigg].
\end{eqnarray}
The modified Hawking temperature of non-stationary black hole depends not only on the properties of the event horizon but also on the mass of the black hole $ M $, cosmological constant $ \Lambda $, charge $ Q $ and on retarded time $ u. $

\section{Discussion}

Firstly, we transform the non static line element of KNdS black hole to static form given in Eq. (5). Applying WKB approximation to Proca equation Eq. (7), Feynman prescription and WKB approximation, the Hawking temperature near the event horizon of KNdS black hole is derived. The Hawking temperature of stationary KNdS black hole is also obtained in Eq. (49) by using deformed Hamilton-Jacobi equation. Since $(1+\sigma m \Theta^2)>1$, the Hawking temperature of stationary black hole is modified. Eq. (49) shows that the Hawking temeprature near the horizon of KNdS black hole will rise due to presence of deformation term $\sigma$ in the Hamilton-Jacobi equation but Refs. [58-62] showed that the quantum gravity effect would prevent the rise of Hawking temperature in black hole. As $ \sigma \rightarrow 0 $, the modified Hawking temperature has been cancelled and original Hawking temperature near the event horizon of KNdS black hole is recovered.

We also investigate the modified Hawking temperature of nonstationary KNdS black hole by using the deformed Hamilton-Jacobi equation, tortoise coordinate transformation and L'Hopital rule. The tortoise coordinate transformation which can be applied on the black hole event horizon [52, 63, 64] is

\begin{eqnarray}
r_*&=&r+\frac{1}{2\kappa_{0}}\ln(r-r_h(u,\theta,\phi))\cr
 u_*&=&u-u_0, \, \,\theta_*= \theta-\theta_0,\,\,\, \phi_*=\phi-\phi_0.
\end{eqnarray}
If we use the tortoise coordinate transformation Eq. (68), the modified surface gravity and the modified Hawking temperature near the event horizon of black hole are 
\begin{eqnarray}
\kappa_{0} =  \frac{r_h\big(1+2\Xi r_{h,u}\big)-M-\frac{2}{3}\Lambda r_h^3-\frac{\Lambda}{3} r_h a^2}{\Big(\Xi (r_h^2+a^2)+\frac{a^2\Xi^{2}\sin^2\theta_0r_{h,u}}{\Delta_\theta}\Big)(1+2r_{h,u})+Z}
\end{eqnarray}
and
\begin{eqnarray}
T_o&=&\frac{1}{2\pi}\frac{r_h\big(1+2\Xi r_{h,u}\big)-M-\frac{2}{3}\Lambda r_h^3-\frac{\Lambda}{3} r_h a^2}{\Big\{\Xi (r_h^2+a^2)+\frac{a^2\Xi^{2}\sin^2\theta_0r_{h,u}}{\Delta_\theta}\Big\}(1+2r_{h,u})+Z},
\end{eqnarray}
respectively.  The chemical potential due to tortoise coordinate transformation (70) is also calculated as 
 \begin{eqnarray}
\omega_0&=&\Big[-eA_0g^{00}r_{h,u}+eA_0g^{01}-(p_\phi+eA_3)g^{03}r_{h,u}-eA_0g^{03}r_{h,\phi}\cr && +(p_\phi+eA_3)g^{13}-g^{22}r_{h,\theta}p_\theta-(p_\phi+eA_3)g^{33}-2\sigma mg^{00}eA_0r_{h,u}\Big]\cr && \times\frac{1}{(g^{01}-g^{00}r_{h,u}-g^{03}r_{h,\phi}+2\sigma mg^{00}r_{h,u})},
\end{eqnarray}
which is same as given in Eq. (65).
 From Eqs. (69) and (64), we get
 \begin{eqnarray}
\kappa=\kappa_{0}+\xi,
\end{eqnarray}
 where
 \begin{eqnarray}
 \kappa_{0}&=&\frac{r_h\big(1+2\Xi r_{h,u}\big)-M-\frac{2}{3}\Lambda r_h^3-\frac{\Lambda}{3} r_h a^2}{\Big\{\Xi (r_h^2+a^2)+\frac{a^2\Xi^{2}\sin^2\theta_0r_{h,u}}{\Delta_\theta}\Big\}(1+2r_{h,u})+Z},\cr
 \xi&=&- \frac{\Delta_\lambda+r_{h,u} \Xi(r_h^2+a^2)+ a\Xi r_{h,\phi}}{r_h\Big[ \big\{\Xi(r_h^2+a^2)+\frac{a^2\Xi^{2}\sin^2\theta_0 r_{h,u}}{\Delta_\theta}\big\}(1+2r_{h,u})+Z\Big]}.
 \end{eqnarray}
 The modified Hawking temperature  at the event horizon is obtained as
 \begin{eqnarray}
 T_h =T_o+\frac{ \xi}{2\pi}.
 \end{eqnarray}
 The correction rate for modified Hawking temperature given in Eq. (67) can be written as
 \begin{eqnarray}
 \alpha &=& \frac{T_{h} - T_{0}}{T_{0}}\cr
 &=& \frac{r_{h}^{-1}\big\{\Delta_{\lambda} + r_{h,u}\Xi(r_{h}^{2} + a^{2}) + a\Xi r_{h,\phi}\big\}}{r_{h}(1 + 2\Xi r_{h,u}) -M-\frac{2}{3}\Lambda r_{h}^{3}-\frac{1}{3}\Lambda r_{h}a^{2}}.
 \end{eqnarray}
 The correction rate is dependent on the parameters: $ M, a, \theta, r_{h,u}, r_{h, \theta}, r_{h, \phi}, \Lambda $ and $ Q $. For stationary black hole, $ r_{h,u}, r_{h,\theta} $ and $ r_{h,\phi} $ tend to zero. Eqs. (68) and (71) give the same Hawking temperature as 
 \begin{eqnarray}
 T_h = T_o = \frac{1}{2\pi} \Bigg[\frac{r_h-M-\frac{2}{3}\Lambda r_h^3-\frac{\Lambda}{3}r_h a^2}{ \Xi (r_h^2+a^2)}\Bigg].
 \end{eqnarray}
 Eqs. (68) and (71) indicate that the Hawking temperatures are modified near the event horizon of nonstationary KNdS black hole. From Eq. (75), as $T_o$ tends to zero, the Hawking temperature $T_h$ will not be zero due to presence of extra term $\xi$.
As $\xi$ tends to zero in Eq. (75), the result is consistent with Eq. (71).  From Eq. (65) and (73), we observe that the chemical potentials are equal due to different tortoise coordinate transformations. It indicates that the tortoise coordinate transformation given in Eq. (55) is more reliable and accurate in the study of modified temperature near the event horizon of KNdS black hole. This shows that the different tortoise coordinate transformations give the different modified Hawking temperatures near the event horizon of KNdS black hole.  If $\sigma$ tends to zero, the different Hawking temperatures derived in Eqs. (28), (50), (69) and (72) by using different methods are equal and consistent with the actual calculation of Hawking temperature given in Eq. (8) near the event horizon of stationary KNdS black hole.

\section{Conclusions}
In this paper, we investigate the tunneling of vector boson particles near the event horizon of static and stationary KNdS black hole using Hamilton-Jacobi ansatz to Proca equation, Feynman Prescription and WKB approximation and corresponding Hawking temperature of the black hole is obtained.

 The modified Hawking temperature near the event horizon of stationary KNdS black hole is also investigated using the modified Dirac equation developed by Kruglov to the deformed Rarita-Schwinger equation describing the fermions. The modified Hawking temperature of KNdS derived from deformed Rarita-Schwinger equation is related to $\sigma m \Theta^2$. It is worth mentioning that the parameter $\Theta$ determines the position of event horizon and the value of $\theta$. The Hawking temperature will rise in the stationary KNdS black hole due to presence of $\sigma$ term in the deformed Hamilton-Jacobi equation. This means that modified Hawking temperature is greater than the original Hawking temperature of stationary KNdS black hole   $ (T > T_{H}) $. As $\sigma$ approaches to zero, the modified Hawking temperature and the original Hawking temperature are equal for the stationary KNdS black hole.

We also studied the modified Hawking temperatures of nonstationary KNdS black hole by taking different tortoise coordinate transformations. The location of event horizon equation on nonstationary space time is obtained by using null surface equation and tortoise coordinate transformation. With the help of event horizon equation, the modified surface gravity and Hawking temperature of nonstationary black hole are obtained.
Using the new tortoise coordinate transformation Eq. (55) in the study of Hawking radiation of black hole, a non zero constant term $\xi$ is appeared in the expression of modified surface gravity and Hawking temperature of nonstationary space time. If $\xi$ is equal to zero, the two modified Hawking temperatures are equal near the event horizon of black hole. Again,  if $\sigma$ and $\xi$ tend to zero, the original Hawking temperature near the event horizon of nonstationary black hole is recovered and is consistent with the earlier results obtained in [57, 63-65]. The different tortoise coordinate transformations give the different modified Hawking temperatures near the event horizon of KNdS black hole but the values of chemical potential are equal. Hence the new tortoise coordinate transformation is more reliable and accurate in the study of modified Hawking temperature of nonstationary black hole. When deformation parameter tends to zero, all the different methods give the same Hawking temperature near the event horizon of stationary KNdS black hole.\\
{\bf Acknowledgement}: The author YKM acknowledges the Council of Scientific and Industrial Research (CSIR), New Delhi for financial assistance.

\end{document}